\documentclass{article}

\usepackage{PRIMEarxiv}

\usepackage[utf8]{inputenc} 
\usepackage[T1]{fontenc}    
\usepackage{hyperref}       
\usepackage{url}            
\usepackage{booktabs}       
\usepackage{amsfonts}       
\usepackage{nicefrac}       
\usepackage{microtype}      
\usepackage{lipsum}
\usepackage{fancyhdr}       
\usepackage{graphicx}       
\usepackage{float}
\usepackage{subcaption}
\graphicspath{{media/}}     

\title{Task-Based Assessment for Neural Networks: Evaluating Undersampled MRI Reconstructions based on Human Observer Signal Detection
}

\author{
  Joshua D. Herman, Rachel E. Roca, Alexandra G. O'Neill, Marcus L. Wong \\
  Mathematics Department \\
  Manhattan College \\
  NY
   \And
  Sajan G. Lingala\\ 
  Roy J. Carver Department of Biomedical Engineering \\
  University of Iowa \\
  Iowa City
  \And
  Angel R. Pineda \\
  Mathematics Department \\
  Manhattan College \\
  NY
}

\begin{document}
\maketitle

\begin{abstract}
Recent research has explored using neural networks to reconstruct undersampled magnetic resonance imaging (MRI) data. Because of the complexity of the artifacts in the reconstructed images, there is a need to develop task-based approaches of image quality.  Common metrics for evaluating image quality like the normalized root mean squared error (NRMSE) and structural similarity (SSIM) are global metrics which average out impact of subtle features in the images.  Using measures of image quality which incorporate a subtle signal for a specific task allow for image quality assessment which locally evaluates the effect of undersampling on a signal. We used a U-Net to reconstruct under-sampled images with 2x, 3x, 4x and 5x fold 1-D undersampling rates. Cross validation was performed for a 500 and a 4000 image training set with both structural similarity (SSIM) and mean squared error (MSE) losses. A two alternative forced choice (2-AFC) observer study was carried out for detecting a subtle signal (small blurred disk) from images with the 4000 image training set.  We found that for both loss functions and training set sizes, the human observer performance on the 2-AFC studies led to a choice of a 2x undersampling but the SSIM and NRMSE led to a choice of a 3x undersampling.  For this task, SSIM and NRMSE led to an overestimate of the achievable undersampling using a U-Net before a steep loss of image quality when compared to the performance of human observers in the detection of a subtle lesion.  
\end{abstract}

\keywords{Task-based assessment \and
Undersampling \and
MRI \and
Neural network \and
2-AFCF}

\section{Introduction}
A major limiting factor in the use of MRI is the long acquisition time.  One of the approaches to decrease imaging time is undersampling the acquired k-space data coupled with constrained reconstruction using total variation or wavelet sparsity priors \cite{Lustig2007, Graff15, PinedaPMB2021}.  Another approach in reconstruction of undersampled MRI data uses data driven priors \cite{Wernick2010,Ravishankar2020, WangDL, Hammernik2018, Aggarwal_2019, ARSHAD2021, Muckley2021} to estimate some of the Fourier information not collected in the data acquisition process.  Some of these approaches incorporate a data agreement term but all introduce prior information in the reconstruction process to compensate for the undersampling.  Since the priors are data driven, the artifacts depend both on the training data and new data being reconstructed since the new data does not exactly match the prior assumptions.  Because of this mismatch, generalization of the results of neural networks being used on new data is an active area of research \cite{KnollAssessment, HUANG2022}.  The artifacts are non-trivial and are often challenging to characterize and predict.  

In a task-based approach to image quality, one needs to specify the task (e.g. detection or estimation), the observer (e.g. human or machine) and the ensemble of images which are being used to quantify the image content \cite{Barrett:90, abbey2001human, Chen2017}.  The image quality depends on these choices.  This task-based approach is particularly useful in evaluating neural network reconstructions because they have artifacts that are non-trivial to characterize.  For other medical imaging modalities, there are recent results that neural networks outside may not lead to improved performance and perhaps lead to decreased performance from a task-based perspective  \cite{Yu2020SPIE, Zhang2021}. 

Evaluation of image quality regarding neural networks in MRI reconstruction typically involves conventional measures of image quality like the normalized root mean squared error (NRMSE) and structural similarity (SSIM) which are simple to compute \cite{Wang2004}.  Another approach uses radiologist's ratings for more realistic tasks \cite{HARPER2021, Recht2020, Lei2022, OBAMA2022} but which require a substantial amount of radiologist's time.  This work explores the space in between using idealized clinical tasks which capture the need of a local and signal dependent version of image quality using non-radiologist observers.  By utilizing these idealized tasks, the objective is to increase the proportion of research which results in clinical impact while minimizing the radiologist's time required.

Preliminary results have been presented at conferences \cite{AccelAbstract, DesignAbstract}.
Here, we assess a U-Net based reconstruction scheme \cite{zbontar2019fastmri} from a task-based perspective of detecting a subtle lesion.  We determine the under-sampling rate that would be chosen based on the proposed human observer studies, and compare it to the undersampling rate that would be chosen based on existing NRMSE and SSIM metrics.  

\section{Methods}
\subsection{Training and Validation Datasets}
In this study, we consider retrospectively undersampling fluid-attenuated inversion recovery (FLAIR) brain datasets from the fastMRI open source dataset \cite{zbontar2019fastmri}.  Coil sensitivity maps were estimated from central 16 k-space lines using the sum of squares approach.  The fully sampled slices were obtained via R=1 SENSE reconstruction \cite{PruessmannSENSE1999} using the BART toolbox \cite{uecker}. We ran five fold cross validation \cite{Hastie2009} using two different sets of images, a 500 image dataset and a 4000 image dataset.  The full data sets (500 and 4000 respectively) were used in the training of the neural networks used for testing the human task performance but the images did not contain the subtle signals used in the human detection experiments.  Those were only included only in the testing set.

\begin{figure}[H]
 \centering
\begin{subfigure}{.45\textwidth}
\centering
  \begin{tabular}{|c|c|}
  \hline
      kx undersampling & effective undersampling\\
      \hline
      2x & 1.90 \\
      \hline
      3x & 2.71 \\
      \hline
      4x & 3.48 \\
      \hline
      5x & 4.16 \\
      \hline
  \end{tabular}
   
  \caption{Effective undersampling factors for each kx undersampling}
  \label{fig:undersampling}
\end{subfigure}
\begin{subfigure}{.45\textwidth}
\centering
  \includegraphics[width=.4\linewidth]{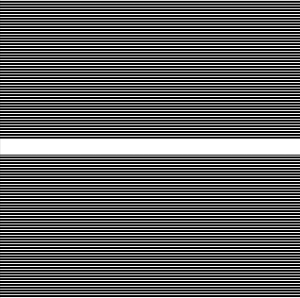}  
  \caption{3x sampling mask}
  \label{fig:mask}
\end{subfigure}
\caption{Undersampling factors and example undersampling mask.  To simulate under-sampled Fourier data, we performed the Fourier transform of the R=1 SENSE reconstruction and multiplied with the undersampling mask.}
\label{fig:undesampling+mask}
\end{figure}

We performed retrospective one-dimensional undersampling on fully sampled images. 
Four 1-D kx undersampling factors: 2x, 3x, 4x, and 5x were implemented. The  undersampling pattern used samples the lowest 16 frequencies (5\% of the data), and sampling every k of the remaining spacial frequencies. Figure \ref{fig:mask} shows a 3x mask.

 While the undersampling factors 2x, 3x, 4x and 5x refer to sampling every 2nd, 3rd, 4th, and 5th of the higher frequency bands respectively, the actual undersampling as the percentage of total frequencies are shown in Figure \ref{fig:undersampling}.  Reconstructed images were normalized in image space to have a minimum pixel value of 0 and maximum of 1.
 
\subsection{Observer Study Testing Dataset}

The images used for the observer studies were based on a 50 image dataset which was not included in either of the training sets.  Four subimages were extracted from each slice \cite{ Oneill2022} to generate a 200 image dataset. The k-space of a small disk (with radius = 0.25 pixels) blurred by a Gaussian kernel ($\sigma =1$ pixel) was added to the k-space of the background images to generate images with a signal in the middle.  The amplitude of the signal was chosen in a pilot study so that the average human performance would be in a range to show differences in undersampling.  The multi-coil k-space data of the  signal was generated by appropriate coil sensitivity weighting and added to the original multi-coil measurements in the test set. For each undersampling factor, this led to a set of 200 background images and 200 images with a signal that were used for the observer studies, but were not used in the training or validation of the neural networks.

\subsection{Neural Network Architecture} 
The neural network design parameters used in this study were chosen as part of a study focusing on picking the preferred dropout and channel numbers for a U-Net using NRMSE, SSIM and human 2-AFC for a fixed undersampling factor of 4x \cite{DesignAbstract}.  In that work, using the SSIM loss function, it was found that when considering SSIM, NRMSE, task-performance and artifacts, the model with 64 channels and 0.1 dropout performed best overall.  Those are the parameters used in this study. 

\begin{figure}[H]
    \centering
    \includegraphics[width=\textwidth]{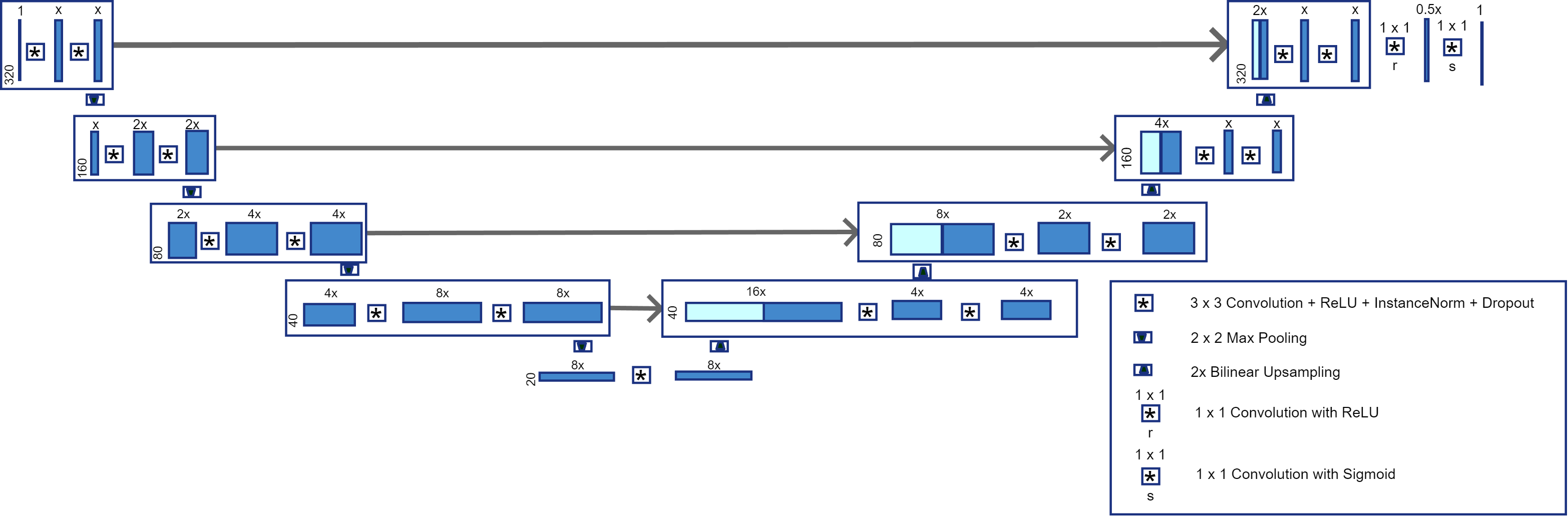}
    \caption{U-Net with $x$ starting channels in the contracting path.  The input of the network is a 320 by 320 undersampled image and the output is the neural network estimate of the 320 by 320 fully sampled image.  The network architecture was based on the baseline network in \cite{zbontar2019fastmri}.  The network used in this study had $x=64$ starting channels and a dropout of 0.1. }
    \label{fig:unet_diagram}
\end{figure}

The U-Net structure consists of two paths and their interconnections. A contracting path is followed by an expanding path, as shown in Figure \ref{fig:unet_diagram}.  The network was based on the baseline network architecture in \cite{zbontar2019fastmri}, except with ReLU and sigmoid activation functions used respectively in the last two layers. We trained our U-Nets with both SSIM and MSE loss functions.

The model starts with a specified number of filters (initial channels), $x$, in its first convolutional block and has 4 pooling layers in the contracting path and 4 upsampling layers in the expanding path. It also contains concatenation blocks that include the output features from each step in the contracting path as input features for each corresponding step in the expanding path.  

\subsection{Training Parameters, Loss Functions and Training Set Sizes} 
Each of the U-Net loss and undersampling schemes were trained using the RMSProp algorithm \cite{Goodfellow-et-al-2016} for 150 epochs and a batch size of 8. The different versions of the U-Net were the SSIM and MSE training loss networks each trained with either a small or large dataset (500 or 4000 training images respectively). We let the SSIM between two images $a$ and $b$ be denoted by $ssim({a,b})$ \cite{Wang2004}.  The ssim metric on which we based our training loss is:
\begin{equation}
SSIM_{METRIC}({y},{\hat{y}})= \frac{1}{m}\sum_{i=1}^{m}{ssim({y_i,\hat{y}_i})},
\end{equation}
where m stands for the number of images, $y$ stands for the array where each entry is a different fully sampled image, while $\hat{y}$ stands for the array where each entry is the corresponding reconstructed image. Since we want our loss to correspond to best picture quality when minimized, we defined our loss to be 
\begin{equation}
SSIM_{LOSS}=1-SSIM_{METRIC}. 
\end{equation}
Our MSE loss was defined by the standard formula for MSE: 
\begin{equation}
MSE_{LOSS}=\frac{1}{m}{\sum_{i=1}^m||{y_i-\hat{y_i}}||_{2}^2}.
\end{equation}

The training of the networks was carried out in an ubuntu Linux workstation with a CPU with 20 threads, 64 GB of DDR4 RAM, and 2 Quadro P5000 16 GB CUDA GPUs.

\subsection{Scoring Metrics}
    We used three scoring metrics to evaluate our reconstructions. Five fold cross validation was used to calculate the mean and standard deviation of SSIM and NRMSE for both the 4000 and 500 training and validation sets.  For the NRMSE we applied the following normalization:
\begin{equation}    
    NRMSE_{METRIC}=\sqrt{\frac{\sum_{i=1}^m||{y_i-\hat{y_i}}||_{2}^2}{\sum_{i=1}^m||y_i||_{2}^2}}.
\end{equation}

\begin{figure}[H]
    \centering
    \includegraphics[width=0.50\textwidth]{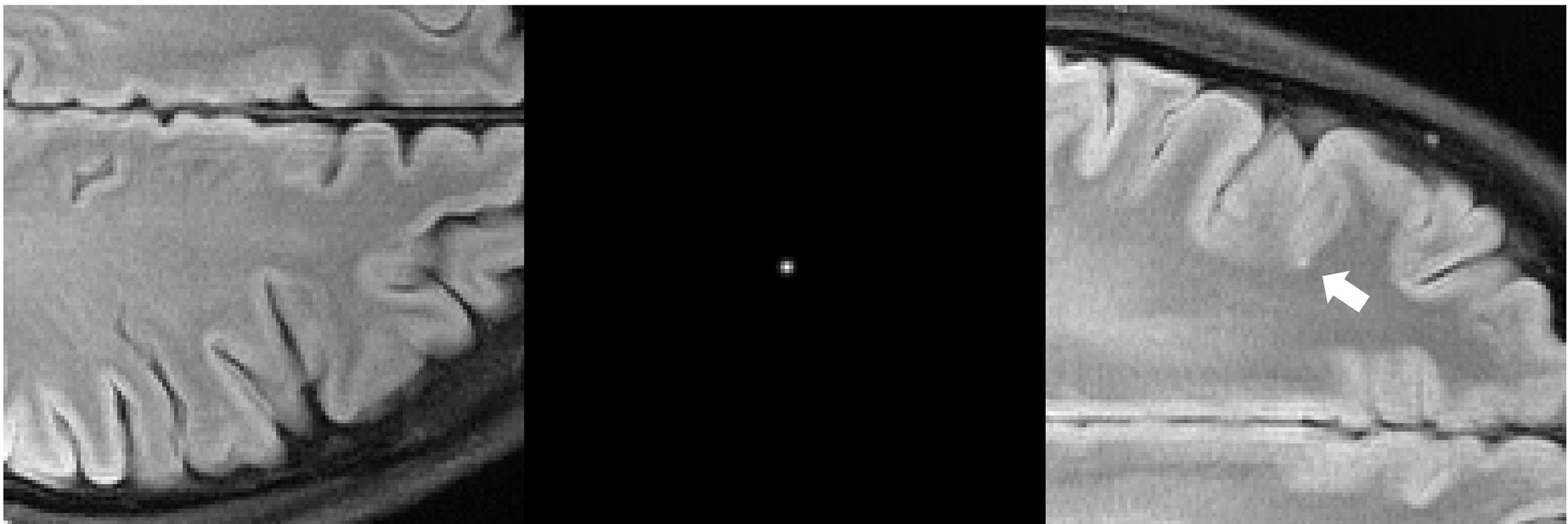}
    \caption{Sample 2-AFC trial for human observer studies using images trained with SSIM loss and large
(4000 image) dataset using 3x undersampling. Image on the right contains signal, with an arrow identifying the location. The image in the middle is the signal. The image on the left is a different background location with no signal. The trial consisted of determining which of the two images contained the signal in the center of the image. The percent correct reported is for 200 trials for each observer and condition.
}
    \label{fig:2-AFC Prompt}
\end{figure}

 The observer performance metric is the average percent correct of the human observers in a two-alternative forced choice (2-AFC) study.  In each 2-AFC trial an observer is presented two images, one containing an artificial lesion (signal image) and another containing no lesion (background image). The task is then correctly selecting the image that contains the lesion.  Since the location of the lesion was always in the center but the backgrounds changed, this is a signal-known-exactly with variable background study \cite{abbey2001human}.  The human observers carried out 200 2-AFC trials for each experimental condition. The percentage of times the person chose correctly is the percent correct. Figure \ref{fig:2-AFC Prompt} is a sample 2-AFC trial.

All the observer studies were done using a Barco MDRC 2321 monitor in a dark room.  The observers were approximately 50 cm from the monitor but a restraint was not used.  The observers took breaks in between trials to avoid fatigue and repeated the first condition until the performance stopped changing as training.

\section{Results \& Discussion}

  Tables \ref{table:imgmetricMSE} and \ref{table:imgmetricSSIM} show results for the 500 image (Small) and 4000 image (Large) 5-fold cross validation SSIM and NRMSE metric scores for the 4 undersampling rates.  The human 2-AFC scores are reported for the 4000 image data-set. Scores are also included for the full reconstruction, which uses fully sampled images and no U-Net reconstruction. The format is mean/standard deviation, and the mean and standard deviation were computed using the five test-train splits for SSIM and NRMSE. Mean and standard deviation for human 2-AFC were computed using the four human observer results. This approach to estimate the standard deviation of observer studies represents an over-estimate of the variability \cite{Gallas2007}. The scores in bold correspond to the preferred undersampling for each metric and the 2-AFC study.

These tables display a small change in mean SSIM and NRMSE from 2x to 3x undersampling, while a larger drop from 3x to 4x. This is true for both training set sizes. Since 3x is the greatest extent you can undersample without a large drop in mean SSIM or NRMSE, 3x is thus the preferred undersampling rate. For human 2-AFC average percent correct, 2x is the preferred amount, since a large drop occurs for mean human 2-AFC from 2x to 3x undersampling. These results suggest that a human needs a more conservative undersampling when compared to NRMSE and SSIM.

 \begin{table}[H]
 \centering
\begin{tabular}{|c|c|c|c|c|c|} 
 \hline
 \textbf{\tiny{Unet 64 channels 0.1 dropout}} & \textbf{SSIM}   & \textbf{SSIM} & \textbf{NRMSE}& \textbf{NRMSE} & \textbf{Human 2-AFC} \\ 
 \hline
 Undersampling & 500 image  & 4000 image & 500 image & 4000 image &4000 image\\
 \hline
 1x & 1 & 1 &0 &0 &{94.500/1.225}\\
\hline
 2x & {0.885/0.004}  & {0.901/0.003} & {0.135/0.019} & {0.114/0.005} &{\bf{92.750/1.299}} \\
 \hline
 \bf{3x} & \bf{0.892/0.006}  & \bf{0.910/0.003} & \bf{0.144/0.020}& \bf{0.108/0.008} &{79.875/4.174}\\
 \hline
 4x & {0.815/0.007}  & {0.825/0.006} & {0.202/0.021} & {0.153/0.007} &{77.625/3.110}\\
 \hline
5x & {0.792/0.012}  & {0.805/0.007} & {0.204/0.027} & {0.163/0.006} &{67.625/2.880}\\
 \hline
\end{tabular}
 \caption{MSE loss U-Net image quality metric results across four undersampling factors and full sampling.}
 \label{table:imgmetricMSE}

\end{table}

   \begin{table}[H]
   \centering
\begin{tabular}{|c|c|c|c|c|c|} 
 \hline
 \textbf{\tiny{Unet 64 channels 0.1 dropout}} & \textbf{SSIM}   & \textbf{SSIM} & \textbf{NRMSE}& \textbf{NRMSE}&\textbf{Human 2-AFC}\\ 
 \hline
 Undersampling & 500 image  & 4000 image & 500 image & 4000 image & 4000 image\\
  \hline
 1x & 1 & 1 &0 &0 &{94.500/1.225}\\
\hline
 2x & {0.914/0.004}  & {0.917/0.002} & {0.118/0.012}& {0.106/0.003} & {\bf{91.625/2.012}}\\
 \hline
 \bf{3x} & \bf{0.912/0.005}  & \bf{0.922/0.001} & \bf{0.136/0.023}& \bf{0.109/0.015} &{77.625/1.746}\\
 \hline
 4x & {0.836/0.008} x & {0.857/0.002} & {0.161/0.009}& {0.148/0.004} &{76.000/3.354}\\
 \hline
5x & {0.814/0.015}  & {0.844/0.003} & {0.178/0.006} & {0.155/0.017}&{64.125/4.762}\\
 \hline
\end{tabular}
 \caption{SSIM loss U-Net image quality metric results across four undersampling factors and full sampling.}
 \label{table:imgmetricSSIM}
\end{table}

\begin{figure}[H]
    \centering
    \includegraphics[width=0.7\textwidth]{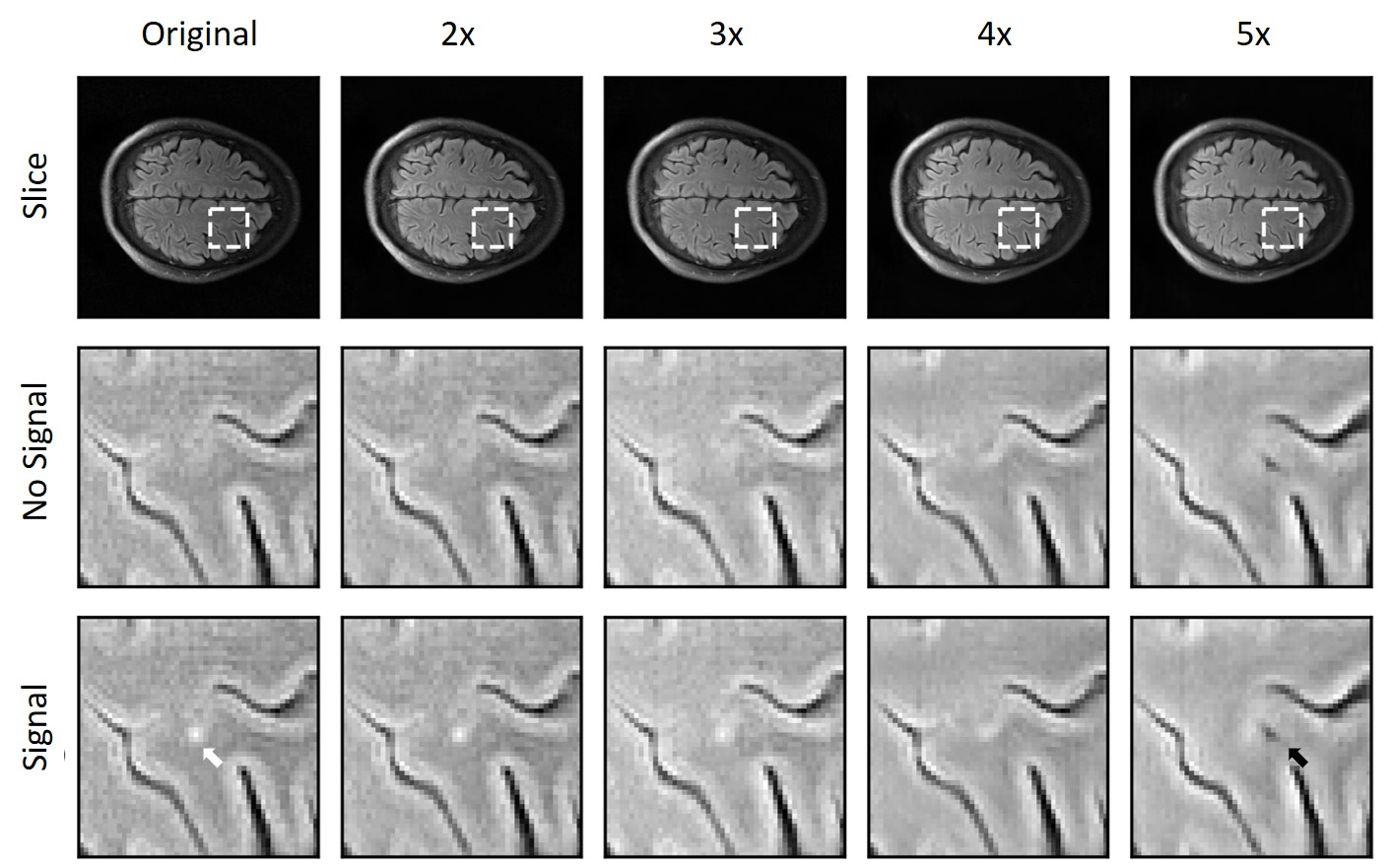}
    \caption{Comparison of undersampled MSE loss images with and without the signal for the large training set. The white arrow points to the signal in the original image and the black arrow to an artifact in the 5x undersampled image.  The signal becomes more faint as the undersampling factor increases. }
    \label{fig:MSEcompare}
\end{figure}

\begin{figure}[H]
    \centering
    \includegraphics[width=0.7\textwidth]{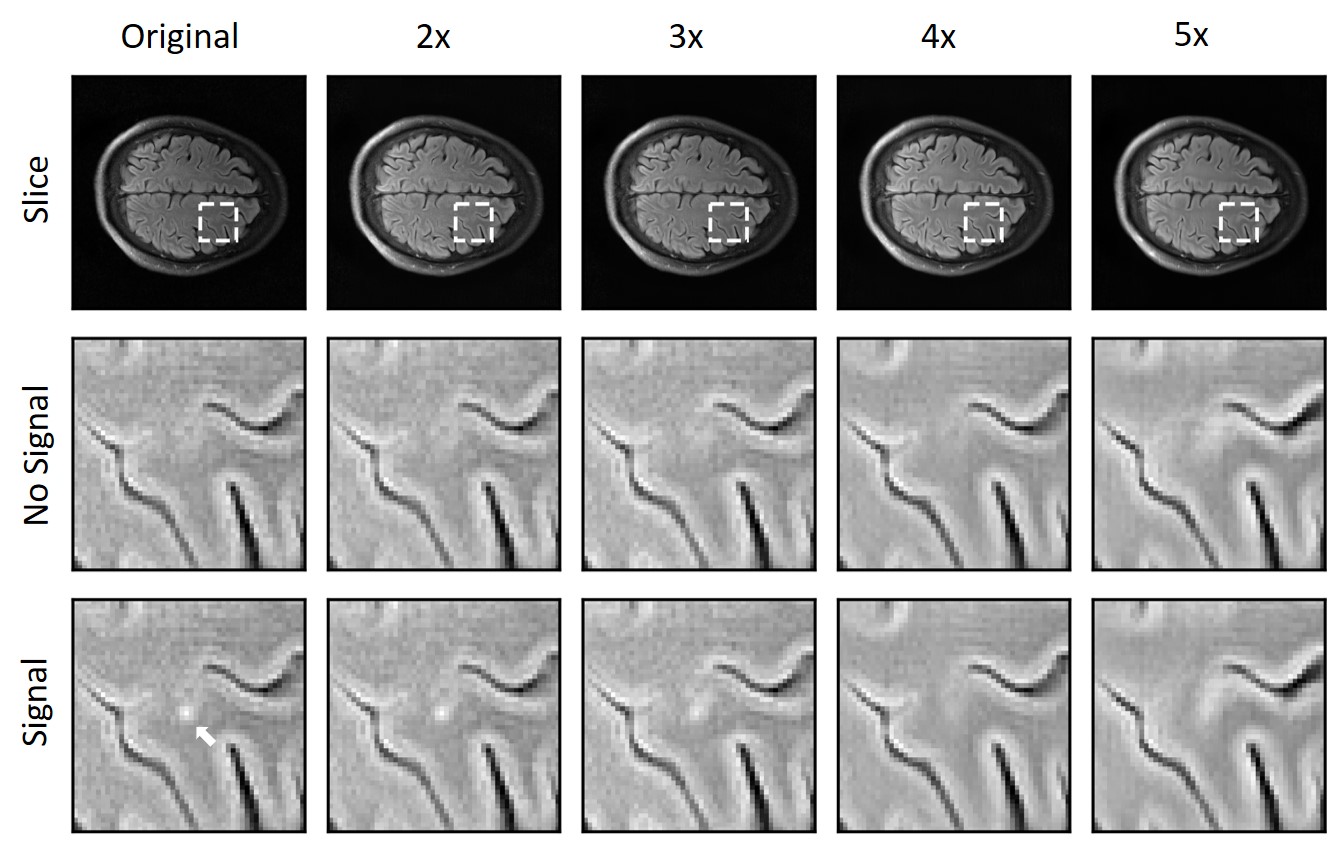}
    \caption{Comparison of undersampled SSIM loss images with and without the signal for the large training set.  The white arrow points to the signal in the original image. This particular image does not have an artifact which resembled anatomy like in the MSE image but similar artifacts were also found for large undersampling factors in the SSIM images.}
    \label{fig:SSIMcompare}
\end{figure}

Subjective evaluation of reconstructed images shown in Figures \ref{fig:MSEcompare} and \ref{fig:SSIMcompare} for the MSE and SSIM loss functions with the large training set shows  that higher undersampling rates affect the visibility of the signal and the artifacts in the images. However, judging the detectability of a signal cannot be done just using a single image because of the significant variability across images. The need for an evaluation using an ensemble of images is one of the justifications of our approach.

Subjective evaluation of the sample images shown in Figures \ref{fig:MSEcompare} and \ref{fig:SSIMcompare} suggests that an undersampling rate of 3x might be acceptable for those images even in a detection task.  Our task performance study is not consistent with that subjective assessment.  This is in part because of the variability of the detection of the signal on the varying backgrounds.  The sample image included an artifact for the 5x undersampling rate for the MSE loss function and not for the SSIM loss function, but those types of artifacts were also present in the SSIM loss images for other subimages.  The artifact looks like a structure and this is sometimes referred to as a hallucination \cite{Muckley2021, Bhadra2021}.   The variability in the use of sample images for assessment and the lack of connection of the NRMSE and SSIM metrics to a task are two of the motivations for the development of task-based approaches.

While current neural network research in image reconstruction mostly focuses on neural networks which enforce data agreement, image domain U-Nets are of interest for analyzing the behavior of neural networks \cite{Hammernik2021} and are being used for specific applications \cite{Jaubert2021}.  In future work we plan to apply a multicoil image implementation of this study which incorporates data agreement using MODL \cite{Aggarwal_2019} instead of a U-Net. We also plan on applying model observers for a 2-AFC task \cite{abbey2001human, Oneill2022} to reduce the number of human observer studies needed for evaluation in this study and incorporate a task where an observer searches for the signal \cite{ Gifford2007, Zhou2020}.  This is part of an overall program to develop task-based assessment in MRI.

\section{Conclusion}
This study introduced human detection performance to evaluate neural networks for undersampled image reconstruction across different undersampling schemes. Scores for two standard image quality metrics, SSIM and NRMSE, and the average human observer performance were calculated for four different undersampling schemes and the fully sampled scheme. This was done using a 4000 image dataset for training and cross validation, with two different loss functions, SSIM and MSE. For NRMSE and SSIM, the 3x undersampling was preferred, while the human observer preferred a more conservative undersampling of 2x.  This work suggests that commonly used metrics for evaluating image quality may over-estimate the amount of undersampling that can be done compared with human detection performance.

\subsection{Code Availability}
The code used for creating, training and running the cross-validation of the UNET as well as for generating the 2-AFC images can be found at: \url{https://github.com/MoMI-Manhattan-College/UNET-2AFC}

\section*{Acknowledgements}
We acknowledge support from the National Institute of Biomedical Imaging and Bioengineering of the National Institute of Health under award number R15-EB029172. We also acknowledge support from the Kakos Center for Scientific Computing and thank Michael Rozycki for his technical support. The authors thank Krishna Nayak, and Dr. Craig Abbey for their helpful insights.

\bibliographystyle{unsrt}  
\bibliography{MRIneuralnetwork}

\end{document}